\documentclass[journal=jacsat,manuscript=article]{achemso}

\usepackage[version=3]{mhchem} 
\usepackage{hyperref}
\usepackage{amssymb}

\DeclareUnicodeCharacter{0308}{\"}
\DeclareUnicodeCharacter{0307}{\"}
\DeclareUnicodeCharacter{030C}{\v}

\author{Ewan R. S. Wallace}
\email{ewan.wallace@roche.com}
\affiliation[Roche]
{Global Informatics, Roche}

\author{Nathan C. Frey}
\affiliation[Prescient]{Prescient Design, Genentech}

\author{Joshua A. Rackers}
\email{rackersj@gene.com}
\affiliation[Prescient]{Prescient Design, Genentech}

\title[An \textsf{achemso} demo]
  {Strain Problems got you in a Twist? \\Try StrainRelief:
  A Quantum-Accurate Tool for Ligand Strain Calculations}

\abbreviations{NNP, SPE, DFT, MAE, ASE, BFGS, LGM, PDB}
\keywords{American Chemical Society, \LaTeX}

\begin{document}


\begin{figure}[htbp]
  \centering
  \includegraphics[width=9cm,height=3.5cm]{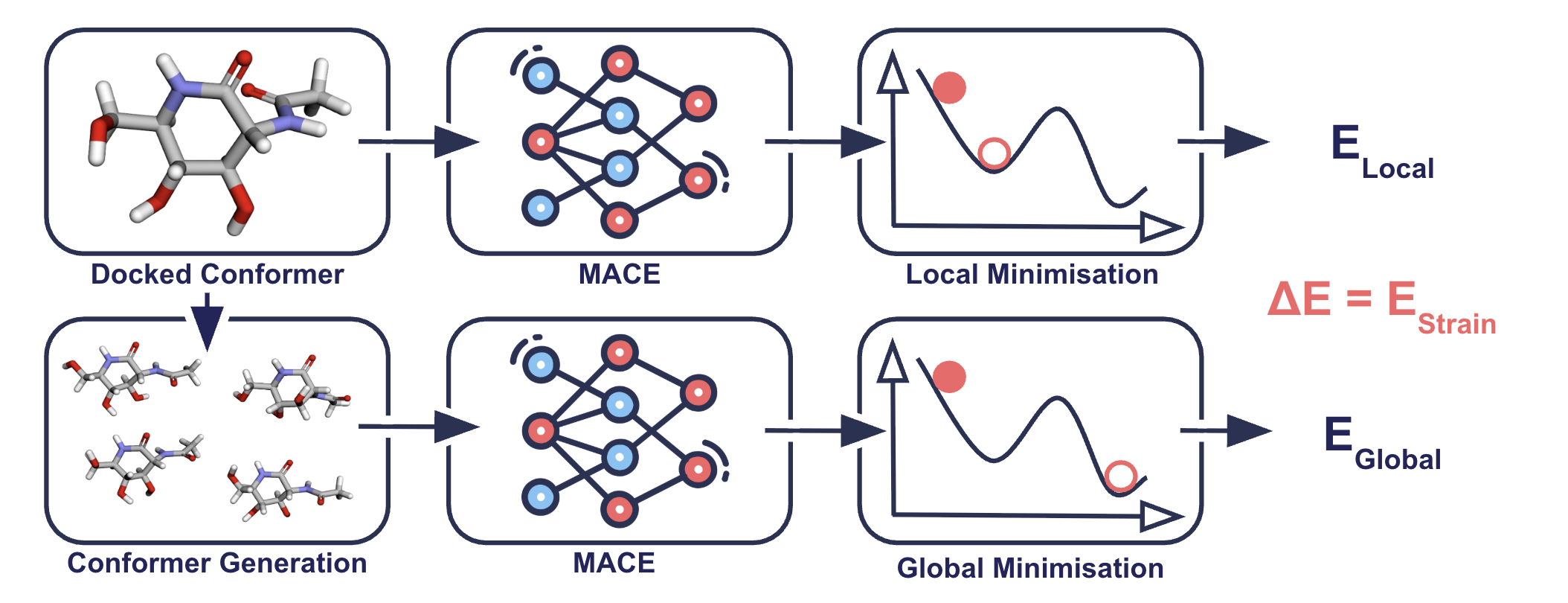}
  \caption{
  TOC Graphic
  }
  \label{fig:toc_graphic}
\end{figure}

\begin{abstract}
Ligand strain energy, the energy difference between the bound and unbound conformations of a ligand, is an important component of structure-based small molecule drug design. A large majority of observed ligands in protein-small molecule co-crystal structures bind in low-strain conformations, making strain energy a useful filter for structure-based drug design. In this work we present a tool for calculating ligand strain with a high accuracy. StrainRelief uses a MACE Neural Network Potential (NNP), trained on a large database of Density Functional Theory (DFT) calculations to estimate ligand strain of neutral molecules with quantum accuracy. We show that this tool estimates strain energy differences relative to DFT to within 1.4 kcal/mol, more accurately than alternative NNPs. These results highlight the utility of NNPs in drug discovery, and provide a useful tool for drug discovery teams.
\end{abstract}

\section{Introduction}

Structure-based drug design has been an incredibly effective strategy for drug discovery since structural biology techniques like x-ray crystallography entered the field in the 1980s.\cite{structuredrugdesign} In this paradigm, new drugs are designed by chemists to fit into an experimentally determined, 3D protein pocket. This “lock and key” style approach intimately relies on accurately predicting and evaluating the pose of any proposed ligand molecule in its pocket. One of the most important metrics in evaluating the quality of a ligand pose is the ligand strain energy.\cite{ligboundconf} Strain energy is defined as the energy difference between the bound and unbound conformations of a ligand (Eqn. \ref{eqn:ligand_strain}), and this metric is valuable because the vast majority of ligand conformations observed in experimentally measured drug-protein complexes have low strain.\cite{ligboundconf} The logic of this is simple. A ligand will only bind to a protein if it is energetically favourable for it to do so. If the conformation required for a ligand to fit into a pocket is too strained, it will not bind. This concept is not just true in the abstract; ligand strain energy has been used as a metric to drive success on real drug discovery projects.\cite{huang2014design, safina2017design} An accurate ligand strain energy filter can have great value for structure-based drug discovery. Here we define ligand strain as an enthalpy, not a free energy, as entropic contributions are not considered. In this paper, we present StrainRelief, a tool that measures ligand strain energy near the accuracy of \textit{ab initio} quantum chemistry.
\begin{equation} \label{eqn:ligand_strain}
E(ligand\ strain) = E(local\ minimum) - E(global\ minimum) \tag{1}\\
\end{equation}
Ligand strain energy has been studied extensively in the literature\cite{strainnicklaus1995conformational,strainallen1996comparison,strainbostrom1998conformational,strainperola2004conformational,strainhao2007torsion,strainfu2011accurate,strainsitzmann2012pdb,strainsellers2017comparison,strainrai2019comprehensive,csd_strain} (an excellent review of previous studies can be found in the introduction of reference \citenum{ligboundconf}). While these studies have some disagreement in how to compute strain and what the energy cut-off should be, they all agree that high-strain molecules should be discarded in drug design workflows. Several tools related to strain energy already exist and can broadly be grouped into two classes: data-driven and physics-based tools. Data-driven tools attempt to estimate ligand strain by analysing distributions of observed small molecule conformations in x-ray crystal structures. These models impose an energy penalty for torsional angles of a ligand that vary too far from the experimentally observed distributions. The most prominent example of this class of models is the model referred to in this paper as CSD, based on the Cambridge Structural Database\cite{csd_strain}. These models suffer from two weaknesses. First, coverage of chemical space in experimental structures can be quite sparse, and second, there is no reference for what the energy penalty for deviation should be. For physics-based tools, models attempt to estimate ligand strain by approximating quantum chemistry. Examples include models based on classical force fields\cite{mm_strain}, and even some using early neural network potentials\cite{ani2}. While these models are based on a rigorous ground truth, they often struggle to accurately approximate quantum chemistry. Previous work has highlighted that force field-based tools can have large errors relative to reference quantum chemistry calculations.\cite{strainsellers2017comparison,strainrai2019comprehensive} Concerningly, there is very little consensus between tools on which conformers are strained. For instance, two commonly used tools\cite{csd_strain, mm_strain} have a Spearman's rank correlation coefficient of only 0.51 between ligand strain predictions (Figure S3).

Clearly, an \textit{ab initio} quantum chemistry strain energy calculator would be a useful strain energy tool, but the computational cost for such a tool would be prohibitive for real-world drug design settings. Fortunately, recent developments in neural network potentials (NNPs) has led to architectures capable of estimating conformational energetics of small organic molecules with extremely high accuracy\cite{maceoff,unke2021machine,wang2024design}. In particular, the equivariant MACE architecture\cite{mace}, which includes higher-order n-body message passing, has shown state-of-the-art performance on Density Functional Theory (DFT) benchmarks for organic molecules.\cite{maceoff} Moreover, NNPs on modern GPU hardware have become fast enough to be useful for the strain energy task.

\begin{figure}[htbp]
  \centering
  \includegraphics[width=0.75\textwidth]{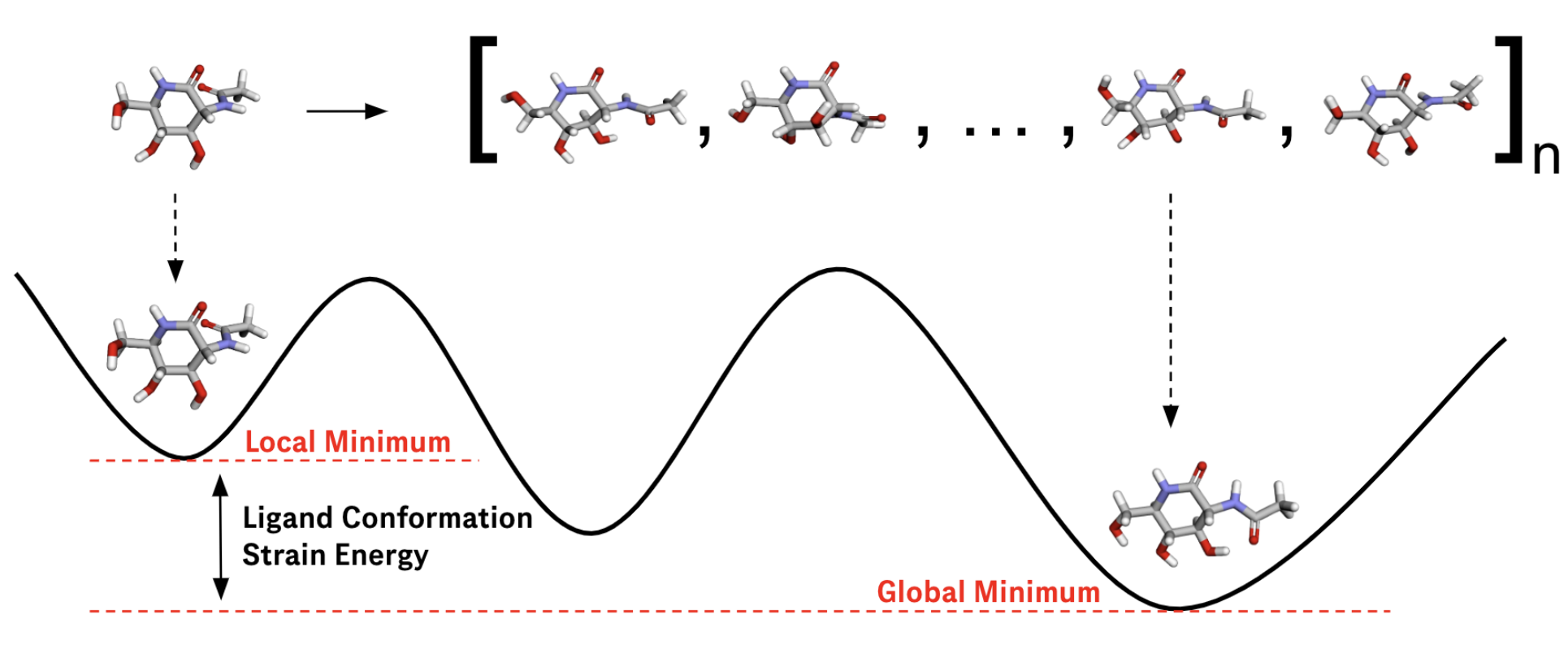}
  \caption{
  The StrainRelief protocol consists of 5 steps: 1. Minimise the docked pose to a local minimum using a loose convergence criteria. 2. Generate n conformers from the docked ligand pose. 3. Minimise all generated conformers. 4. Evaluate energies of all minima 5. Calculate ligand strain (Eqn. \ref{eqn:ligand_strain} \& \ref{eqn:global_min}).
  }
  \label{fig:protocol}
\end{figure}

In this work, we train a MACE NNP on a large database of DFT calculations and use that model to calculate strain energies of drug-like molecules. We show that this model has excellent accuracy relative to single point energy (SPE) DFT calculations on relevant molecules outside of its training set. We show that a simple protocol for gas phase strain energy (Figure \ref{fig:protocol}) is sufficient to capture the observed distribution of strains in real-world drug-like molecules. And, moreover, we show that by combining the trained MACE model with the simple protocol, we get a straightforward calculator that classifies docked poses as strained or unstrained with high accuracy. 

All of this is combined into the StrainRelief tool, which we make publicly available at \url{github.com/prescient-design/StrainRelief}.

\begin{equation} \label{eqn:global_min}
E(global\ minimum) \approx min[E(conformer\ 0\ minimum),\ ...\ ,\ E(conformer\ n\ minimum)]\tag{2}
\end{equation}

\section{Results}

A ligand strain tool consists of a model and protocol pairing. The model describes the potential energy surface. The protocol defines the steps taken to calculate the ligand strain.

Here, we propose a set of 3 criteria essential for the development of an accurate ligand strain tool:
\begin{enumerate}
    \item \textbf{Accurate Energy Evaluation:} The model must comprehensively characterise the potential energy surface, accurately evaluating the energies and forces of both high- and low-energy conformers.
    \item \textbf{Global Minimum Conformer Approximation:} The model and protocol pairing must successfully approximate global minimum conformations.
    \item \textbf{Enrichment of Active Samples:} The model and protocol pairing must effectively enrich the pool of active samples (compounds showing potential therapeutic value through biological screening assays) by systematically excluding ligands that exhibit significant strain.
\end{enumerate}

By meeting these criteria, the proposed ligand strain tool will effectively identify and exclude ligand structures with high strains, enhancing the drug development pipeline.

\subsection{Accurate Energy Evaluation}

Predicting ligand strain is a particularly challenging task due to the need for a model to accurately characterise the entire potential energy surface. This characterisation includes not only rare, high energy strained conformers produced by docking software but also low energy global minima for energy evaluations, as well as all possible conformers in between to ensure stable dynamics.

\begin{figure}[htbp]
  \centering
  \includegraphics[width=1.0\textwidth]{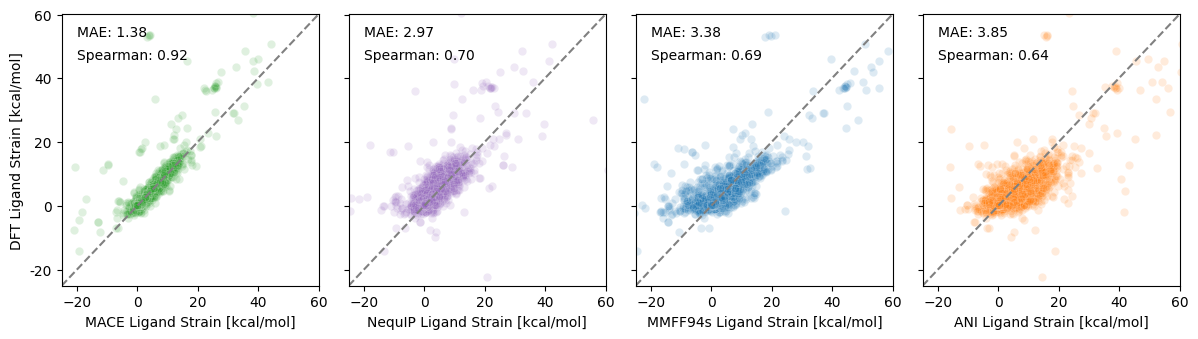}
  \caption{
  LigBoundConf 2.0 ligand strains calculated by evaluating local and global minimum conformers using different force fields (MACE, NequIP, MMFF94s, ANI). Predictions are compared to DFT. Note: the same conformers are evaluated in each plot - determined using xTB+CREST (see Methods - LigBoundConf 2.0).
  }
  \label{fig:xtb_conf_energies}
\end{figure}

In this study, a selection of NNPs, including MACE\cite{mace}, ANI\cite{ani1,ani2}, and NequIP\cite{nequip}, along with a commonly used empirical potential, MMFF94s,\cite{mmff_rdkit} were evaluated on the local and global minimum conformers of our own curated neutral subset of the LigBoundConf dataset\cite{ligboundconf}, LigBoundConf 2.0 (see Methods - LigBoundConf 2.0) (Figure \ref{fig:xtb_conf_energies}). All NNPs were trained on neutral molecules from SPICE 2\cite{spice} - a dataset curated to encapsulate protein-ligand interactions with maximally diverse conformers. Conformers are evaluated with DFT at the $\omega$B97M-D3(BJ)\cite{functional1, functional2}/def2-TZVPPD\cite{basis_set1, basis_set2} level of theory. This level of theory is chosen for its balance of speed and accuracy on quantum chemistry benchmarks.\cite{dft_study} To unlock significantly higher accuracy a double hybrid functional or quadruple zeta basis set would be needed, both of which are prohibitively expensive for a dataset of this size. Among the models tested, MACE demonstrated superior performance, exhibiting a significantly lower mean absolute error (MAE). MACE is roughly equally accurate in evaluating bound and global minimum conformers with MAEs of SPE predictions of 1.85 and 1.72 kcal/mol respectively. Crucially, a Spearman’s rank correlation of 0.92 indicates that MACE effectively identifies the strain of ligands relative to one another - an important quality for a filter. Despite its overall strong performance, MACE tends to perform poorly in certain cases, with predictions deviating by as much as 10 or 20 kcal/mol. Zwitterionic ligands are particularly problematic, exhibiting an average MAE that is 4.7 kcal/mol higher than other ligands. These zwitterionic ligands account for 62\% of cases with errors above 10 kcal/mol and 57\% of cases with errors above 20 kcal/mol. This can be attributed to the fact that charged molecules in the gas phase tend to form unphysical shapes, and all charged molecules were excluded from the SPICE 2 training data.

MACE generalises well to LigBoundConf 2.0, with only 16 ligands being common between this and the SPICE 2\cite{spice} training set. Excluding common ligands, a further 2 ligands are considered to be highly structurally similar to any molecules in the SPICE 2\cite{spice} training set, determined by a minimum Tanimoto difference $\leqslant 0.15$ (Figure S1).\cite{tanimoto_threshold} MACE's exceptional performance is consistent with findings from recent literature on this NNP architecture\cite{maceoff, citing_mace1, citing_mace2, citing_mace3, citing_mace4}. MACE achieves a MAE of 1.32 kcal/mol (0.068 kcal/mol per atom) on the SPICE 2 PubChem test set when predicting absolute energies. This result aligns with models trained with identical hyperparameters in MACE-OFF23\cite{maceoff}, which report a MAE of 0.063 kcal/mol per atom on SPICE 1 PubChem. Furthermore, MACE aligns closely with the MACE-OFF23 (S) model on the TorsionNet dataset and shows considerable improvements over other models presented in this study (Figure S2). Due to its excellent performance, as show in Figure \ref{fig:xtb_conf_energies}, MACE has been selected as the standard model for StrainRelief and is utilised throughout this paper, unless specified otherwise.

\begin{figure}[htbp]
  \centering
  \includegraphics[width=0.6\textwidth]{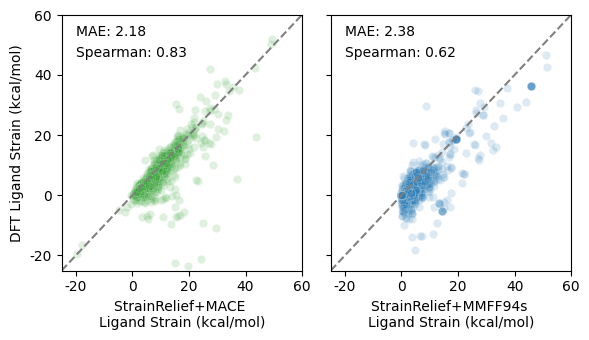}
  \caption{
  StrainRelief is run on LigBoundConf 2.0 ``raw” conformers. Local and global minima generated by the protocol are evaluated with DFT. Different models are used in the plot: MACE (left) and MMFF94s (right). MACE local minimum and global minimum MAEs are 3.56 and 4.83 kcal/mol respectively. Note: as the conformers are generated using model-specific minimisations, different conformers are evaluated in each plot.
  }
  \label{fig:dynamics_mae_from_raw}
\end{figure}

In addition to evaluating given conformers, StrainRelief seeks to accurately approximate minima. We ran the StrainRelief protocol with our trained MACE model on the LigBoundConf 2.0 dataset to obtain both local and global minima (Figure \ref{fig:dynamics_mae_from_raw}). The resulting local minimum structures have an average RMS displacement of 0.23 Å (95th percentile = 0.54 Å) relative to the ``raw” structures. We then evaluated the energy difference of those minima with DFT to assess the accuracy of our model+protocol. We did the same experiment with the MMFF94s model as a baseline. All minimisations were run with the Broyden–Fletcher–Goldfarb–Shanno\cite{bfgs} (BFGS) algorithm as implemented in Atomic Simulation Environment\cite{ase} (ASE) using their MACE calculator. 

As expected, errors increase when evaluating conformers generated by StrainRelief’s own dynamics, with local minima and global minima having SPE MAEs of 3.56 and 4.83 kcal/mol respectively. The SPICE training data is pulled from high temperature molecular dynamics and so, by design, it does not sample global minima. This leads to worse model performance on these conformers and the model could potentially be improved by fine-tuning on a set of low energy conformations. MMFF94s shows significant improvement in MAE over it's performance in Figure \ref{fig:xtb_conf_energies}, however it's poor ranking of ligands makes it an unwise choice of force field. Additionally, DFT verifies that MACE identifies a lower energy, and therefore better, global minima in 86\% of cases when compared with MMFF94s.

Despite the increase in MACE's error relative to Figure \ref{fig:xtb_conf_energies}, its Spearman's rank correlation remains relatively high, indicating that errors arising from conformer generation are largely systematic instead of random. This ensures that the ranking of ligand strains is still largely correct and that StrainRelief is a useful filter.

\subsection{Global Minimum Conformer Approximation}

For 36.5\% of molecules in LigBoundConf 2.0, DFT confirms that the global minimum identified by StrainRelief is lower energy than the one found by xTB+CREST, despite CREST enumerating 50 conformers for it's global minimum search in comparison to StrainRelief's 20. Additionally, a tolerance of only 0.51 kcal/mol is required to increase this to 50\%. This demonstrates that although simple, when paired with an accurate model, StrainRelief's protocol accurately approximates global minima in a significant proportion of cases, and at a rate 400 times faster than xTB+CREST. This drastic increase in speed facilitates high throughput workflows.

\subsection{Enrichment of Active Samples}

The ultimate goal of a ligand strain tool is to filter out strained, inactive compounds from a set of docked poses. Because strained compounds are highly likely to be inactive, this should enrich the proportion of active drug candidates. Unfortunately, negative outcomes are rarely recorded at scale in drug discovery, leading to a distinct lack of publicly available negative or decoy ligand strain data to test such tools. Due to this lack of public data, we demonstrate StrainRelief's ability to identify strained molecules and enrich active drug candidates using LigBoundConf 2.0.

\begin{figure}[htbp]
  \centering
  \includegraphics[width=0.5\textwidth]{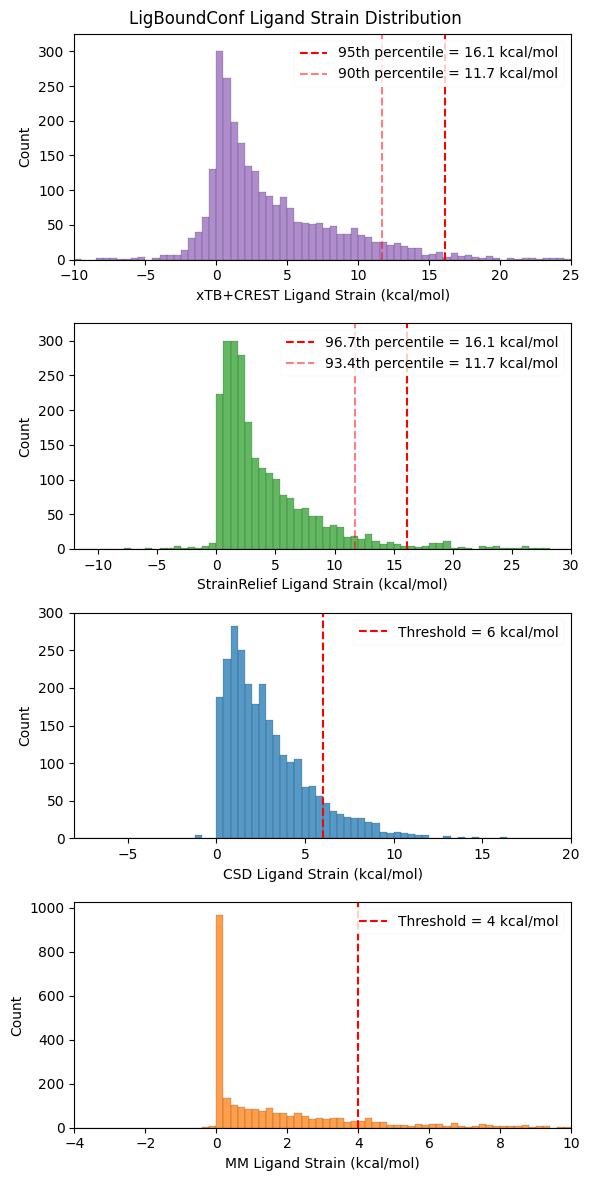}
  \caption{
  Distributions of ligand strains for different protocol-model pairings. All pairings start from LigBoundConf 2.0's ``bound" configurations (as an approximation of docked poses). MM and CSD thresholds are taken from suggestions in their respective papers.\cite{mm_strain, csd_strain} The 90th and 95th percentiles are taken from the xTB+CREST workflow and imposed upon StrainRelief, with new percentiles calculated. Thresholds define a cut-off above which molecules should be considered strained when evaluated with a given tool.
  }
  \label{fig:protocol_histograms}
\end{figure}

The majority of ligand conformations observed in experimentally measured drug-protein complexes exhibit low strain\cite{ligboundconf}. This observation is corroborated by all four protocols depicted in Figure \ref{fig:protocol_histograms}, where, despite different energy ranges and methodologies, the distributions consistently show a preference for lower strains. However, despite the similarity in distribution, there is minimal agreement between protocol's ranking of ligand strains. Spearman's rank correlation values are calculated to be below 0.51 between all pairs of tools, except for xTB+CREST and StrainRelief, which show a correlation of 0.69 (Figure S3). This discrepancy likely arises from fundamental differences in the protocol and model pairing used by each tool.

The MM strain\cite{mm_strain} tool is a chemalot command-line workflow (sdfMMConfAnalysis) and is the most similar of the two literature methods to StrainRelief. It calculates a local minimum and performs a global minimum search by enumerating conformers. Two key differences are that both an implicit solvent model and an empirical force field (MMFF94s) are used. Figure \ref{fig:protocol_histograms} demonstrates that, for 749 of LigBoundConf 2.0's 2726 ligands, the search for the global minimum failed to find anything lower energy than the local minimum and so predicts a strain of 0 kcal/mol.

The CSD strain tool\cite{csd_strain} adopts a data-driven approach by summing individual torsions. The CSD is used to sample from an energy distribution for all possible dihedral angles. Each dihedral in a molecule is listed, and their energies are evaluated from the relevant distribution. The final strain is the sum of all molecular dihedral torsions.

\begin{figure}[htbp]
  \centering
  \includegraphics[width=1.0\textwidth]{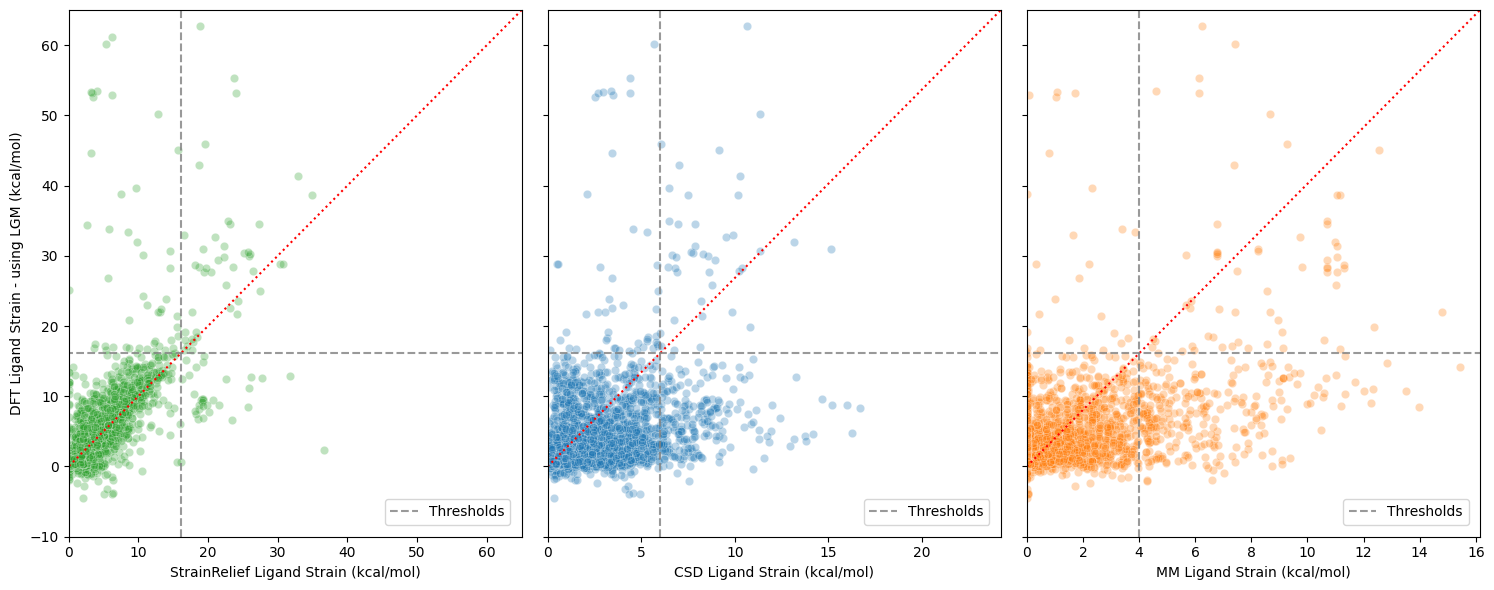}
  \caption{
  The performance of different protocols in identifying strained LigBoundConf 2.0 ligands (i.e., above the 95th percentile). The y-axis uses an ``lowest global minimum" (Eqn. \ref{eqn:lowest_global_minimum}) for ligand strain (Eqn. \ref{eqn:lgm_strain}). Protocols are run from LigBoundConf 2.0's ``bound" poses as an approximation of docked conformers. Evaluation metrics for each plot are summarised in Table \ref{tbl:protcol_comparison}.
  }
  \label{fig:scatterplot_filter}
\end{figure}

To evaluate the performance of each protocol as a filter, we need to establish a ``ground truth" strain. Since StrainRelief accurately identifies a lower energy global minimum than xTB+CREST in 36.5\% of cases, a lowest global minimum (LGM) is used (Eqn. \ref{eqn:lowest_global_minimum}). This LGM is determined by choosing the lowest energy of 3 global minimum conformers suggested by three tools and evaluated with DFT. The three tools are CREST, StrainRelief-N20, and StrainRelief-N100, accounting for 56, 27, 17\% of global minima, respectively (N here refers to the number of conformers requested - see SI). In Figure \ref{fig:scatterplot_filter}, the y-axis represents the DFT energy difference between this LGM and the xTB local minimum (Eqn. \ref{eqn:lgm_strain}). It should be noted that MM strain global minima cannot be included in the LGM as they use an implicit solvent, and CSD does not compute minima.
\begin{align} \label{eqn:lowest_global_minimum}
\lefteqn{Lowest\ Global\ Minimum\ (LGM) =} \tag{3}\\
& & min[E_{\text{CREST}}(global\ min),\ E_{\text{StrainRelief-N20}}(global\ min),\ E_{\text{StrainRelief-N100}}(global\ min)] \tag{3}
\end{align}
\begin{equation} \label{eqn:lgm_strain}
E(ligand\ strain) = E_{\text{xTB}}(local\ min) - LGM \tag{4} \\
\end{equation}

\begin{figure}[htbp]
  \centering
  \includegraphics[width=0.75\textwidth]{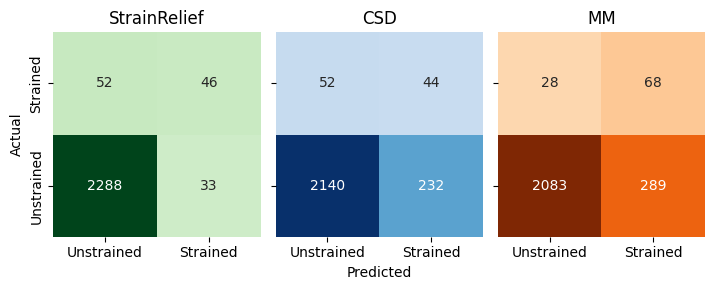}
  \caption{
  Confusion matrices for the StrainRelief, CSD and MM protocols in Figure \ref{fig:scatterplot_filter}.
  }
  \label{fig:confusion_matrices}
\end{figure}

\begin{table}
  \caption{Filter performance metrics for each tool, as plotted in Figure \ref{fig:scatterplot_filter}.}
  \label{tbl:protcol_comparison}
  \begin{tabular}{lllll}
    \hline
    Protocol  &  Spearman's R & Precision & Recall & F1 Score  \\
    \hline
    StrainRelief  & \textbf{0.770} & \textbf{0.582} & 0.469          & \textbf{0.520} \\
    CSD           & 0.358          & 0.159          & 0.458          & 0.237 \\
    MM            & 0.480          & 0.190          & \textbf{0.708} & 0.300 \\
    \hline
  \end{tabular}
\end{table}

StrainRelief can be evaluated for two primary purposes: as a binary strain filter and as a relative ligand strain predictor. StrainRelief significantly outperforms the CSD and MM tools in ranking of ligand strains with respect to quantum chemistry, as evidenced by Spearman's rank correlation values in Table \ref{tbl:protcol_comparison}. Both CSD and MM perform very poorly in this task, and are largely uncorrelated with the DFT LGM strains (Figure \ref{fig:scatterplot_filter}).

As a binary strain filter StrainRelief also outperforms other tools, achieving an F1 score (the harmonic mean of precision and recall) of 0.520 (when defining strained molecules as positive results) (Table \ref{tbl:protcol_comparison}). The threshold used for StrainRelief and DFT calculations is the 95th percentile (16.1 kcal/mol) of LigBoundConf 2.0 ligand strains. The most critical metrics in this context are precision and the number of false positives (unstrained molecules incorrectly identified as strained) — areas where StrainRelief excels relative to other tools. Minimising false positives is crucial in drug discovery processes to avoid discarding potentially efficacious molecules. While false negatives can still proceed to subsequent filtering stages, false positives are immediately dismissed from the pipeline. This means that high precision and low false negative rates are integral to maintaining a high-quality dataset for further testing. Although the MM correctly identifies a high number of strained ligands, its higher false positive rate undermines its effectiveness as a filter.

All tools underestimate very high ligand strains (above 40 kcal/mol), which exclusively have LGMs derived from xTB. This is a possible area for future data augmentation when training subsequent models.

\section{Methods}

\subsection{LigBoundConf 2.0} \label{sec:LigBoundConf}

To evaluate the accuracy of our NNP model and strain energy protocol, we constructed a reference dataset from the benchmark LigBoundConf database. LigBoundConf is a set of high-quality drug-like bound ligand structures in the Protein Data Bank (PDB). The original set contains three sets of structures: ``raw” structures taken directly from the PDB, ``bound” structures, obtained by performing a local minimisation of the ligand in the presence of the protein pocket with an empirical force field, and ``global minimum” structures, obtained by finding the lowest energy conformation of the ligand with an empirical force field and implicit water solvation model. Because we adopt a definition of strain energy that does not include protein interaction or solvation, we chose to run our own set of calculations on the LigBoundConf dataset. Starting from the ``raw” ligand structures, we created our own set of ``bound”, and ``global minimum” structures. This new version of the dataset we call ``LigBoundConf 2.0”.

To obtain the ``bound” structures we sought to relieve any artificial high-energy artifacts from the deposited PDB structures. We did this by running a local minimisation with the semi-empirical quantum chemistry method GFN2-xTB using the “crude” convergence criteria. This ensured that artifacts were removed without significantly altering the bound pose of the ligand. The resulting minimised structures have an average RMS displacement of 0.20 Å (95th percentile = 0.47 Å) relative to the ``raw” structures. We then performed SPE DFT calculations at the $\omega$B97M-D3(BJ)\cite{functional1, functional2}/def2-TZVPPD\cite{basis_set1, basis_set2} level of theory.

To obtain the ``global minimum” structures, we used the CREST enumeration method with 50 conformers, which uses the same GFN2-xTB energy function to search for low energy gas phase structures. We took the lowest energy structure as the ``global minimum” conformation and performed a SPE DFT calculation at the same $\omega$B97M-D3(BJ)\cite{functional1, functional2}/def2-TZVPPD\cite{basis_set1, basis_set2} level of theory. 

Note that this protocol is only strictly valid for neutral molecules. Charged molecules are liable to undergo electrostatic collapse in the gas phase (Figure S4). Moreover, including a solvation model in strain energy is troublesome; bound ligands are typically at least partially buried in protein pockets, so including solvation in a reference calculation for the bound structure is unlikely to be a useful model.

The LigBoundConf 2.0 neutral dataset ended up with 2726 ligands. The strain energy of these ligand’s difference in DFT energies between the ``bound” and ``global minimum" conformations is plotted in the first histogram of Figure \ref{fig:protocol_histograms}. This clearly shows that drug-like ligands are likely to bind with low strain energy, in agreement with the original LigBoundConf dataset. The 95th percentile of the distribution lies at 16.1 kcal/mol, which we adopted as our strained vs. unstrained classification cutoff. We use this cutoff as the classification mechanism for the subsequent experiments in this work.

\subsection{StrainRelief Protocol}

The protocol used in StrainRelief is designed to be simple, fast and model agnostic - all that is needed to apply a new force field is to write an ASE\cite{ase} calculator wrapper. The protocol consists of 5 steps:

\begin{enumerate}
    \item Minimise the docked pose with a loose convergence criteria to give a local minimum.
    \item Generate 20 conformers from the docked ligand pose.
    \item Minimise the generated conformers (and the original docked pose) with a stricter convergence criteria.
    \item Evaluate the energy of all conformers and choose the lowest energy as an approximation of the global minimum.
    \item Calculate ligand strain (Eqn. \ref{eqn:ligand_strain} \& \ref{eqn:global_min}).
\end{enumerate}

\textbf{Conformer Generation:} 20 additional conformers are requested from the given docked pose using RDKit’s conformer generation with default ETKDG (v2) parameters and a UFF force field. If not given, bonds are also determined using RDKit prior to this. Due to RMSD pruning, the requested number of conformers represents an upper bound on the number of conformers actually generated. For the LigBoundConf 2.0 dataset, requesting 20 conformers results in an average of 16.5 returned.

\textbf{Minimisation:} All minimisations are done using an adapted version of the BFGS\cite{bfgs} algorithm as implemented in ASE,\cite{ase} with default values for $max\ step$ (0.2 Å) and $alpha$ (70.0). Minimisation is killed if the number of steps exceeds 250 or if during any step the $max(force) > 250\ eV/$Å. ASE calculator wrappers were written for both MACE and RDKit’s MMFF94 force fields with the intention that other calculators could be easily written in a similar fashion. \textbf{i) Local Minimisation:} The given docked pose is minimised with a loose convergence criteria ($max(forces) < 0.50\ eV/$Å). Determining a local minimum with your chosen force field is necessary to ``clean-up" ligand structures. Docked  poses can often contain artificially high energy artifacts, such as slight deviations in ideal bond lengths or angles. \textbf{ii) Global Minimisation:} All generated conformers (and the docked pose) are minimised with the default BFGS convergence criteria of $max(forces) < 0.05\ eV/$Å.

\textbf{Energy Evaluation:} If a different force field is being used for energy evaluations and minimisations (this is not recommended), perform SPE evaluations on the local minimum and all potential global minima.

\textbf{Strain Calculation:} 
\begin{equation} 
E(ligand\ strain) = E(local\ minimum) - E(global\ minimum) \tag{1}\\
\end{equation}
\begin{equation}
E(global\ minimum) \approx min[E(conformer\ 0\ minimum),\ ...\ ,\ E(conformer\ n\ minimum)]\tag{2}
\end{equation}
Ligand strains $< 0$ (i.e. where $E(local\ minimum) < E(global\ minimum)$) are set to zero. Computational efficiency scales linearly with conformers generated (not conformers requested). On a GPU, StrainRelief calculations average 44 seconds per molecule (with 20 conformers). 

Note: in 10\% of molecules in LigBoundConf 2.0 StrainRelief failed to determine a strain value because either the local minimum or all global minima failed to converge within the required number of steps. In the case of negative predicted strains, 21 ligands were assigned a strain of zero.

\subsection{The SPICE Dataset\cite{spice}}

All NNPs in this paper were trained on the SPICE 2.0.1 dataset. SPICE is a quantum chemistry dataset specifically designed for training potentials relevant to simulating drug-like small molecules and their interactions with proteins. SPICE conformers are generated by running molecular dynamics simulations and sampling maximally different conformations. These conformers are then evaluated at the $\omega$B97M-D3(BJ)\cite{functional1, functional2}/def2-TZVPPD\cite{basis_set1, basis_set2} level of theory. SPICE contains a number of  subsets, with PubChem being the largest, accounting for 70\% of all conformers. SPICE is a useful training set for a ligand strain model because it samples many and varied high- and low-energy conformers across the drug-like small molecule space.
For our training, we excluded the ion pairs subset, removed all metal-containing compounds (leaving only H, C, N, O, F, P, S, Cl, Br, and I) and removed all molecules with a non-zero formal charge. This leaves 1,834,342 conformers. We trained on an 80-10-10 train-test-validation split, stratified by subset.

\subsection{Models}

\subsubsection{MACE\cite{mace}}

MACE is a high order equivariant messaging passing architecture that scales well up to 4 body interactions. For our model, we used the same hyperparameters as in MACE-OFF23 (S)\cite{maceoff}: chemical channels = 96, max L = 0, correlation = 3, cutoff radius = 4.5 Å. Although larger MACE models tend to be more accurate, they also perform inference more slowly. As our intended purpose for our tool is high throughput screening we opted for a smaller model.

Models were trained using PyTorch and the training scripts in the mace repository\cite{mace_repo} with identical hyperparameters to those reported for MACE-OFF23(S)\cite{maceoff}. The model was trained for 190 epochs, with a loss function combining a linear combination of energy and force predictions. Following the recommendations in the MACE-OFF23 paper, training was initially weighted towards forces with a learning rate of 0.01. It is then switched to weighting towards energies for the final 35 training epochs\cite{maceoff} and stochastic weight averaging (SWA) was used with a decreased learning rate of 0.00025. The Adam optimiser with Amsgrad was used and the exponential moving average of the weights was taken in each training step. The exact hyperparameters used are available in the StrainRelief GitHub repository (\url{examples/mace_training_script.sh}).

\subsubsection{ANI\cite{ani1, ani2}}
ANI is an equivariant deep neural network that uses a highly modified version of the Behler and Parrinello symmetry functions to build single-atom atomic environment vectors as molecular representations. The ANI-2x architecture was trained using the open-source PhysicsML repository\cite{physicsml}.

\subsubsection{NequIP\cite{nequip}}
NequIP is an early equivariant graph neural network iteratomic potential. As recommended in the original paper when training small-molecule systems, we used 5 iteration blocks, a learning rate of 0.01 and a batch size of 5. Training for NequIP was conducted using the open-source PhysicsML repository\cite{physicsml}.

\section{Conclusion}

Ligand strain energy is a useful metric for filtering out unrealistic candidates in drug discovery. Although studies differ on precisely how best to measure the strain energy, all agree that there is an upper limit to how much strain a ligand can tolerate and still bind to its target. In this work we show that a strain energy calculator built on the assumption that quantum chemistry provides a reasonable ground truth yields an accurate method for estimating ligand strain. This NNP-based calculator classifies strained v.s. unstrained molecules with very high accuracy.

There are clear limitations to this approach. Because it is gas-phase, StrainRelief does not handle charged molecules. Solving this problem in the context of a quantum strain energy calculator will require careful consideration, as calculating strain energy in the context of aqueous solvation is a poor approximation of protein pocket environment. Additionally, there is much more that goes into the free energy of drug-protein binding than strain energy. Strain may be a computationally inexpensive way to remove unrealistic bound conformations, but this is only one of many barriers in identifying high-affinity drug candidates.

Despite these limitations, the StrainRelief tool presented here is both accurate and useful. This shows that NNPs can approximate quantum chemistry accurately enough to be useful in a real-world application context. More importantly, it shows that this model can be used as an accurate empirical filter in drug discovery campaigns.

\subsection{Author's Contributions}

EW trained all NNPs used, wrote the code for the StrainRelief repository, and ran all experiments. JR led the project, providing insight and guidance at each step, and also computed the LigBoundConf 2.0 dataset. EW and JR jointly wrote the paper. NCF provided guidance on code development and helpful comments on the paper.

\subsection{Data and Software Availability}

The StrainRelief \href{github.com/prescient-design/StrainRelief}{repository} is available at \url{github.com/prescient-design/StrainRelief}. This contains all the code needed to run the tool, as well as example scripts and notebooks. The 3 sets of \href{huggingface.co/datasets/erwallace/LigBoundConf2.0}{LigBoundConf 2.0} conformers (``raw", ``bound" and ``global minimum"), their DFT energies, and the StrainRelief tool's results when run on the ``bound" conformers can all be found at \url{huggingface.co/datasets/erwallace/LigBoundConf2.0}.

\subsection{Conflicts of Interest}

We have no conflicts of interest to disclose.

\begin{acknowledgement}

The authors would like to thank the MACE-OFF23 authors for the use of their recalculated TorsionNet DFT values at the SPICE level of theory (Figure S2).

\end{acknowledgement}

\begin{suppinfo}

Additional information detailing the choice of N for conformer generation, the distribution of Tanimoto differences between training and tests sets, additional test set metrics, and examples of charged structures undergoing electrostatic collapse (PDF). 

\end{suppinfo}


\bibliography{references}

\end{document}